\documentclass[preprint]{jpsj2}

\renewcommand{\theenumi}{( \( \alph{enumi} \) )}
\def\runtitle{
 A New Approach to Stochastic State selections in Quantum Spin Systems 
}
\def\runauthor{Tomo  \textsc{Munehisa} and Yasuko \textsc{Munehisa}}

\title{%
 A New Approach to Stochastic State Selections in Quantum Spin Systems 
}

\author{%
Tomo  \textsc{Munehisa} and
Yasuko \textsc{Munehisa}
}

\inst{%
 Faculty of Engineering, Univ. of Yamanashi, Kofu 400-8511
}

\recdate{\today}

\abst{%

We propose a new type of Monte Carlo approach 
in numerical studies of quantum systems.
Introducing a probability function which determines whether a state in  
the vector space survives or not, we can evaluate expectation values of
powers of the Hamiltonian from a small portion of the full vector
space. This method is free from the negative 
sign problem because it is not based on importance sampling techniques.
In this paper we describe our method and, in order to examine how
effective it is, present numerical results on
the $4\times 4$, $6\times 6$ and $8\times 8$ Heisenberg spin one-half 
model. The results indicate that we can perform useful evaluations with 
limited computer resources. An attempt to estimate the lowest energy
eigenvalue is also stated.
}

\kword{%

quantum spin, large size, numerical calculation, Monte Carlo,
eigenvalue
}

\begin{document}
\sloppy
\maketitle

\section{Introduction}

A great many methods have been investigated so far to numerically calculate
various quantities in quantum spin systems. They are classified into two
categories, the exact diagonalization \cite{dagetal} and 
the Monte Carlo \cite{book1,book2} approaches.
Both of them provide us very useful ways to study quantum spin systems 
but, as was repeatedly reported, the former is difficult to apply to large 
size systems and the latter often suffers from the negative sign
problem.
In this paper we introduce a new method based on the power
method\cite{book2,pwm} in order to evaluate eigenvalues of these systems. 
Our purpose is to show that one can effectively investigate the ground
state by means of a stochastic selection of states.

The method we propose here is a kind of Monte Carlo approach, 
where stochastic variables play an important role. It, however,  
differs much from the conventional quantum Monte Carlo methods which
employ random walks or importance samplings\cite{book2}.
In our method random variables are used to reduce the number of states
of the vector space which is huge for most systems of large sizes.
To this end we consider a new type of the probability function 
to which we refer as {\em on-off probability} function in this paper. 
Since it ``{\em switches off}'' many states in the
vector space we can calculate approximate expectation values of powers of 
the Hamiltonian from a small number of the ``$on$'' states.
Repeating this process we can obtain averaged values which are very
close to the exact values. 
Physical quantities such as the
energy eigenvalue of the system would then be estimated from these 
expectation values with some additional assumptions.
Throughout this paper we use the two-dimensional Heisenberg spin 
one-half model as a concrete example in numerical calculations.
 
In the next section we define the on-off probability function and 
explain its analytical properties.  
We then show, in a simple instance using the exact ground state for 
the $4\times4 $ Heisenberg spin system,  
how we make use of this probability function when we calculate a 
quantity concerning to the vector space. Numerical results are presented
in order to demonstrate that it nicely works.
Section 3 is to describe our approach in a quantum system whose 
Hamiltonian is $\hat H$. We define random choice matrices and present a way to 
stochastically calculate expectation values of 
$\langle \psi \mid \hat H ^L \mid \psi \rangle$ with them.  
Two cases of $\mid \psi \rangle$ which draw our interest are then discussed.
One is that the state is the exact eigenstate of  $\hat H$, 
$\mid \psi \rangle = \mid \psi _{\rm E} \rangle$, and another is an
approximate of $\mid \psi _{\rm E} \rangle$ which we denote as 
$\mid \psi _{\rm A} \rangle$. In section 4 we present our numerical results on
the Heisenberg spin systems. Results on a $4\times4$ lattice, where it
is easy to obtain $\mid \psi _{\rm E} \rangle$, are given 
in detail so that we can make a close inspection of the method.
Then results on the $6\times6$ and $8\times8$ lattices are presented 
together with the description of the state $\mid \psi _{\rm A} \rangle$ we
generated. We also show several assumptions we employed to estimate the 
energy eigenvalues for these lattice sizes. The values we present in
this section for the $6\times6$ and $8\times8$ lattices are compatible 
with the ones reported in the preceding studies with different 
approaches\cite{square}. 
The final section is devoted to summary and comments.

\section{On-off Probability Function}

Let us introduce a probability function which is given by
\begin{eqnarray}
 P(\eta) = \frac{1}{a}\delta(\eta -a) +(1- \frac{1}{a})\delta(\eta),
\label{on-off}
\end{eqnarray}
where $a$ is a constant which is greater than or equal to 1.
In this paper we refer it as the $on$-$off$ $probability$ $function$ 
because the variable $\eta$ takes only two values, $a$ ($on$) or 0
($off$), with the probability $1/a$ and $1-1/a$, respectively.
Clearly
\begin{eqnarray}
   \langle \! \langle  1 \rangle \! \rangle  & \equiv & \int_0^\infty P(\eta) d \eta =  1 , \\  
\langle \! \langle   \eta \rangle \! \rangle &\equiv & \int_0^\infty \eta  P(\eta) d \eta = 1,
\label{eta1}
 \\
  \langle \! \langle   \eta^2 \rangle \! \rangle  &\equiv & \int_0^\infty \eta^2  P(\eta) d \eta = a ,
\label{eta2}
\end{eqnarray}
where $\langle \! \langle  \rangle \! \rangle$ denotes the
statistical average.
In order to demonstrate the role of this probability function
we consider, as a simple example, the sum of absolute values of all (non-zero) 
coefficients in a state 
$\mid \psi \rangle =$ $ \sum \mid i \rangle c_i$ ($c_i \neq 0$) 
of the two-dimensional Heisenberg spin model, 
\begin{eqnarray}
  S \equiv \sum_{i=1}^n |c_i| ,
\label{sdef}
\end{eqnarray}
and try to evaluate it by summing up {\em less} number of $|c_i|$'s.
For this purpose we calculate the following sum 
\begin{eqnarray}
  S_{\{ \eta \}} \equiv \sum_{i=1}^n |c_i| \eta_i ,
\label{seta} 
\end{eqnarray}
where $\{ \eta \}$ denotes a set of random variables $\{ \eta_1, \eta_2,
\cdots, \eta_n \}$ generated by the following on-off probability
functions  
\begin{eqnarray}
  P_i(\eta_i) & = &
\frac{1}{a_i}\delta(\eta_i -a_i) +(1- \frac{1}{a_i})\delta(\eta_i) ,
\label{pi}
\end{eqnarray}
$a_i$ being $max(1,\epsilon/| c_i|)$
for a given constant $\epsilon $.
It is easy to see that 
\begin{eqnarray*}
 \langle \! \langle  S_{\{ \eta \}}  \rangle \! \rangle   = \sum_{i=1}^n |c_i| \langle \! \langle \eta_i\rangle \! \rangle  = S .
\end{eqnarray*}
The variance of $ S_{\{ \eta \}} $ is given by
\begin{eqnarray}
\sigma_{\rm S}^2  \equiv \langle \! \langle  S_{\{ \eta \}}^2  \rangle \! \rangle  -  \langle \! \langle  S_{\{ \eta \}}  \rangle \! \rangle ^2 
 = \sum_{  | c_i | < \epsilon } (\epsilon | c_i |-|c_i|^2 ) 
\label{sigdef}
\end{eqnarray}
where $\langle \! \langle \eta_i \eta_j \rangle \! \rangle  
= \delta_{ij} \ \langle \! \langle \eta_i^2 \rangle \! \rangle +(1-\delta_{ij})\langle \! \langle \eta_i \rangle \! \rangle \langle \! \langle \eta_j \rangle \! \rangle 
=\delta_{ij}a_i+(1-\delta_{ij}) $ is used.

\noindent
Let $N_i = 1 (0)$ if $\eta_i \neq 0 (=0)$ and $N = \sum N_i$. Then we
know by $\langle \! \langle  N \rangle \! \rangle $ 
how many non-zero $\eta_i$'s appear in (\ref{seta}). Since     
\begin{eqnarray*}
 P_{N_i}( N_i) &= &
 \frac{1}{a_i}\delta( N_i -1)+(1-\frac{1}{a_i})\delta( N_i), \\
\langle \! \langle  N_i  \rangle \! \rangle &\equiv& \int_0^{\infty}  N_i  P_{N_i}( N_i) d  N_i = \frac{1}{a_i},
\end{eqnarray*}
we obtain
\begin{eqnarray}
\langle \! \langle  N \rangle \! \rangle  = \langle \! \langle  \sum_{i=1}^n N_i\rangle \! \rangle  = 
\sum_{  | c_i | \geq \epsilon } 1
      +\sum_{  | c_i | < \epsilon } \frac{ | c_i | }{\epsilon} .
\label{ndef}
\end{eqnarray}

In numerical work we substitute the average over $n_{\rm smpl}$ randomly generated 
${\{ \eta \}}$'s for the statistical average 
$\langle \! \langle  \cdots \rangle \! \rangle $, namely we measure  
\begin{eqnarray}
\langle \! \langle  S_{\{ \eta \}} \rangle \! \rangle _{\rm smpl} \equiv 
\frac{1}{n_{\rm smpl}} \sum_{k=1}^{n_{\rm smpl}} S_{\{ \eta \}_k},
\label{wem1}
\end{eqnarray} 
where $\{ \eta \}_k$ denotes the $k$-th set of the generated random variables 
$\{ \eta \}=\{ \eta_1, \eta_2, \cdots, \eta_n \} $. 
The variance is then given by
\begin{eqnarray}
\rho_{\rm S}^2 \equiv
\langle \! \langle  S_{\{ \eta \}}^2\rangle \! \rangle _{\rm smpl}
-\langle \! \langle  S_{\{ \eta \}}\rangle \! \rangle _{\rm smpl}  ^2 . 
\label{wem2}
\end{eqnarray}  
As the statistical error we employ 
\begin{eqnarray}
Er \equiv 2\sqrt{
\frac{ \ \rho_{\rm S}^2}{n_{\rm smpl}}} 
\ . 
\label{wem3}
\end{eqnarray}
Note that, according to the Chebyshev inequality in statistics, 
\begin{eqnarray*}
\langle \! \langle  S_{\{ \eta \}} \rangle \! \rangle _{\rm smpl} - Er \ \leq \ \langle \! \langle  S_{\{ \eta \}} \rangle \! \rangle  \  \leq \ 
\langle \! \langle  S_{\{ \eta \}} \rangle \! \rangle _{\rm smpl} + Er 
\end{eqnarray*}
in 75 \% confidence level provided that $\sigma_{\rm S}^2$ can be replaced by 
$\rho_{\rm S}^2$\cite{foot1}. 

The results for the exact ground state of the $4\times 4$ Heisenberg spin
model, which has 12870 non-zero coefficients, are shown in Table I. In 
this case it is straightforward to calculate 
$S (= \langle \! \langle  S_{\{ \eta \}} \rangle \! \rangle 
 \ )$, $\sigma_{\rm S}^2 $ and $\langle \! \langle  N \rangle \! \rangle $ 
from (\ref{sdef}), (\ref{sigdef})
and (\ref{ndef}) since we know all $c_i$ of the state. In Table I we see that  
$\langle \! \langle  S_{\{ \eta \}} \rangle \! \rangle _{\rm smpl}$,
$\rho_{\rm S}^2$ and $\langle \! \langle  N \rangle \! \rangle _{\rm smpl}$ 
calculated with $n_{\rm smpl} = 10^4$ 
are in good agreement with them. We also see that 
$\epsilon=0.2$ is enough to estimate $S$ in the precision of $ 0.1 \% $.
In other words we can estimate $S$ in the precision of $ 0.1 \% $ from 
$only$ 367 of 12870 non-zero coefficients of the state. 
 
\section{Powers of Hamiltonian }
In this section we discuss on the expectation values of the $L$-th 
power of the Hamiltonian $\hat H$.
First we make a brief comment on the power method to evaluate 
the maximum eigenvalue of $\hat{H}$. 
Suppose the eigenstates of $\hat{H}$ are $\mid \psi _i \rangle$, 
$\hat H \mid \psi_i \rangle= \mid \psi_i \rangle E_i $ $(i=0,\cdots, N_{\rm V}-1
)$ where $N_{\rm V}$ is the size of the full vector space, 
and $\mid E_0 \mid > \mid E_{i} \mid$ for $i > 0$. 
In the power method one repeatedly operates $\hat H$ to a trial state 
$\mid \psi  \rangle$, which is expressed as 
\begin{eqnarray} 
\mid \psi \rangle = \sum_{i=0}^{N_{\rm V}-1} \mid \psi_i \rangle b_i , 
\label{psitry}
\end{eqnarray}
so that one obtains for sufficiently large $L$
\begin{eqnarray} 
 \hat H^L \mid \psi \rangle = \sum_{i=0}^{N_{\rm V} -1 } \mid \psi_i \rangle 
b_i E_i^L \ \sim \ \  \mid \psi_0 \rangle b_0 E_0^L
\label{hlll} 
\end{eqnarray}
and
\begin{eqnarray}
  \langle \psi \mid \hat H^L \mid \psi \rangle \sim b_0^2 E_0^L .
\label{hlevll} 
\end{eqnarray}
Note that from the expectation values $\langle \psi \mid \hat H^L \mid
\psi \rangle$ one can acquire information relative to the energy eigenvalue. 
For large size systems, however, it soon becomes
impracticable to calculate $ \hat H^L \mid \psi \rangle$ or
$\langle \psi \mid \hat H^L \mid \psi \rangle$ 
because one can not keep the huge number of coefficients for 
$ \hat H^L \mid \psi \rangle$ in the full vector
space. One therefore has to do with truncated vector spaces.

Now we describe our formulation. Let us start with a state 
\begin{eqnarray} 
\mid \psi \rangle = \sum_{i=1}^{N_{\rm V}} \mid i \rangle c_i , 
\label{psitry2}
\end{eqnarray}
where $\{ \mid i \rangle\}$ denotes an arbitrary basis of the full
vector space.
In our method we stochastically find a limited vector space where we can 
calculate the expectation values
$ E(L) \equiv \langle \psi \mid \hat H^L \mid \psi \rangle$ for large $L$. 
For this purpose we introduce an 
$N_{\rm V} \times N_{\rm V}$ diagonal matrix $M_{ \{ \eta \}}$,
which we call {\em random choice matrix} hereafter,
\begin{eqnarray}
M_{ \{ \eta \}} \equiv 
\left( \begin{array}{cccc} \eta_1 & 0 & \cdots & 0 \\
0 & \eta_2 & \cdots & 0  \\
\cdots & \cdots & \cdots & \cdots \\
 0 & 0 & \cdots & \eta_{N_{\rm V}} \end{array}  
\right) \ ,
\end{eqnarray} 
with random variables $\eta_i$ determined according to (\ref{pi}).
When $c_i = 0$ we employ $\epsilon / \delta$ as $a_i$ in (\ref{pi}), where 
the parameter $\delta (< \epsilon)$ is a positive constant independent of $i$. 

Using independent random choice matrices 
$M_{\{ \eta ^{(m)}\}} = {\rm diag.}\{ \eta_1^{(m)}, \eta_2^{(m)},\cdots, 
\eta_{N_{\rm V}}^{(m)}\}$ $(m = 1, 2, \cdots$, $L+1)$ we define  
\begin{eqnarray}
  E_{\{\eta\}}(L) \equiv \langle \psi |M_{\{ \eta^{(L+1)}\}} \hat H 
M_{\{ \eta ^{(L)}\}} \hat H M_{\{ \eta ^{(L-1)}\}} \cdots \hat H 
M_{\{ \eta ^{(1)}\}} | \psi \rangle .
\label{eldef}
\end{eqnarray}
Here we use the coefficient $c_i$ of the $initial$ wave function 
$\mid \psi \rangle$ in the on-off probability function (\ref{pi}) 
for $any$ $m$ . 
In order to explain the essential point of the method in numerical
studies, let $n_{\rm b}(m)$ and $n_{\rm a}(m)$ the number of non-zero components 
of the state 
$\hat H M_{\{ \eta ^{(m-1)}\}} \cdots \hat H 
M_{\{ \eta ^{(1)}\}} | \psi \rangle $ and  
$M_{\{ \eta ^{(m)}\}} \hat H M_{\{ \eta ^{(m-1)}\}} \cdots \hat H 
M_{\{ \eta ^{(1)}\}} | \psi \rangle $, respectively. 
We first keep $n_{\rm b}(1)$ non-zero components of $\mid \psi
\rangle$. Then we operate $M_{\{ \eta ^{(1)}\}}$ to $\mid \psi \rangle$.
Since many of the random variables $\eta_i^{(1)}$ $(i=1,2, \cdots, N_{\rm V})$
are zero, the state $M_{\{ \eta ^{(1)}\}} \mid \psi \rangle$ has much
less non-zero components ($i.e.$ $n_{\rm a}(1) \ll n_{\rm b}(1)$). The number of
non-zero components increases when we next
operate $\hat H$ to $M_{\{ \eta ^{(1)}\}} \mid \psi \rangle$ 
($n_{\rm b}(2) \ > \ n_{\rm a}(1)$), but it
drastically decreases after operating  $M_{\{ \eta ^{(2)}\}}$ to the state 
($n_{\rm a}(2) \ll n_{\rm b}(2)$). Choosing the parameter $\epsilon$ in the on-off
probability functions we can repeat the operation 
up to $L$ within the range of our computer facilities. 

From (\ref{eldef}) we are led to, 
with $h_{ij} = \langle i | \hat H | j \rangle$, 
\begin{eqnarray}  
\langle \! \langle E_{\{\eta \}}(L) \rangle \! \rangle & =&
\langle \! \langle  \sum_i  \sum_j \cdots \sum_k \sum_l
c_i \eta_i^{(1)} h_{ij} \eta_j^{(2)} \cdots
 \eta_k^{(L)}  h_{kl} \eta_l^{(L+1)} c_l  \rangle \! \rangle \nonumber \\
& =&\sum_i  \sum_j \cdots \sum_k \sum_l c_i h_{ij}\cdots h_{kl} c_l  \  
\langle \! \langle  \eta_i^{(1)}  \eta_j^{(2)}  \cdots   
\eta_k^{(L)} \eta_l^{(L+1)}\rangle \! \rangle \nonumber \\
& =&\sum_i  \sum_j \cdots \sum_k \sum_l c_i h_{ij}\cdots h_{kl} c_l
=  \langle \psi | \hat H ^L | \psi \rangle = E(L) 
\label{eetal}
\end{eqnarray}
because 
$\langle \! \langle  \eta_i^{(1)}  \eta_j^{(2)}  \cdots   \eta_k^{(L)}
\eta_l^{(L+1)}\rangle \! \rangle = \langle \! \langle  \eta_i^{(1)}
\rangle \! \rangle \langle \! \langle  \eta_j^{(2)}\rangle \! \rangle  \cdots  
\langle \! \langle \eta_k^{(L)}\rangle \! \rangle  \langle \! \langle \eta_l^{(L+1)}\rangle \! \rangle 
= 1$, which is guaranteed by the fact that $\eta_i^{(m)}$ and
$\eta_j^{(m')}$ are mutually independent for $m \ne m'$. 
Henceforth we abbreviate $\langle \! \langle \eta_i^{(m)}\rangle \! \rangle $
and $\langle \! \langle [ \eta_i^{(m)} ]^2 \rangle \! \rangle $
as $\langle \! \langle \eta_i\rangle \! \rangle $ and $\langle \! \langle \eta_i^2 \rangle \! \rangle $ because they do not depend on $m$. 
It should be noted that the variances of $\eta_i^{(m)}$ are the
same for all $m$.

The variance of $E_{\{\eta \}}(L)$ is

\newpage
 
\begin{eqnarray}
\sigma_{\{\eta \}}^2(L) &\equiv&  
\langle \! \langle E_{\{\eta \}}(L)^2 \rangle \! \rangle -
\langle \! \langle E_{\{\eta \}}(L) \rangle \! \rangle \ ^2 \nonumber \\
&=&  \sum_{i,i'}  \sum_{j,j'} \cdots \sum_{k,k'} \sum_{l,l'}
c_i h_{ij}\cdots h_{kl}c_l \cdot c_{i'} h_{i'j'}\cdots h_{k'l'} c_{l'} 
 \nonumber \\
& &\times \ 
\langle \! \langle  \eta_i^{(1)}\eta_{i'}^{(1)}\rangle \! \rangle \langle \! \langle  \eta_j^{(2)}\eta_{j'}^{(2)}\rangle \! \rangle   \cdots 
\langle \! \langle  \eta_k^{(L)}\eta_{k'}^{(L)}\rangle \! \rangle \langle \! \langle  \eta_l^{(L+1)}\eta_{l'}^{(L+1)}\rangle \! \rangle 
\ - E(L)^2 
\nonumber \\
&=&  \sum_{i,i'}  \sum_{j,j'} \cdots \sum_{k,k'} \sum_{l,l'}
c_i h_{ij}\cdots h_{kl}c_l \cdot c_{i'} h_{i'j'}\cdots h_{k'l'} c_{l'}
\cdot \{1 + \delta_{ii'}( \ \langle \! \langle  \eta_i^2 \rangle \! \rangle -1)\}
\nonumber \\
& & \times \ 
\{1 + \delta_{jj'}( \ \langle \! \langle  \eta_j^2 \rangle \! \rangle -1)\}
 \cdots
\{1 + \delta_{kk'}( \ \langle \! \langle  \eta_k^2 \rangle \! \rangle -1)\}
\{1 + \delta_{ll'}( \ \langle \! \langle  \eta_l^2 \rangle \! \rangle -1)\} \nonumber \\ 
& -&  E(L)^2,
\label{s2etal}
\end{eqnarray} 
where we use  
$\langle \! \langle  \eta_i^{(m)}\eta_{i'}^{(m)}\rangle \! \rangle = (1-\delta_{ii'})\langle \! \langle  \eta_i\rangle \! \rangle \langle \! \langle  \eta_{i'}\rangle \! \rangle 
+\delta_{ii'} \ \langle \! \langle  \eta_i^2 \rangle \! \rangle 
=1+\delta_{ii'}( \ \langle \! \langle  \eta_i^2 \rangle \! \rangle -1)$.

Let us add a few expressions which are useful to check our numerical
values presented in the next section. They are obtained from (\ref{eetal}) and 
(\ref{s2etal}) when the state $ \mid \psi \rangle$ is 
an exact eigenstate $ \mid \psi _{\rm E} \rangle = \sum \mid \psi_i \rangle f_{i} \
(f_i \neq 0)$ of $\hat H$, namely when $\hat H |
\psi _{\rm E} \rangle =| \psi _{\rm E} \rangle E$ or $\sum_j h_{ij} f_{j} = E
f_{i}$.  
\begin{eqnarray}
\langle \! \langle E_{\rm E{\{\eta \}}}(L)  \rangle \! \rangle  &=&  E_{\rm E}(L) = E^L,
\\
\sigma_{\rm E{\{\eta \}}}^2(1)  &=& 
2E^2 \sum_i f_{i}^4( \ \langle \! \langle  \eta_i^2 \rangle \! \rangle -1)
\nonumber \\
&+& \sum_{i,j}  f_{i}^2 f_{j}^2 (h_{ij})^2 
( \ \langle \! \langle  \eta_i^2 \rangle \! \rangle -1)
( \ \langle \! \langle  \eta_j^2 \rangle \! \rangle -1), 
\label{sig1} 
\\
\sigma_{\rm E{\{\eta \}}}^2(2)  &=& 
3E^4 \sum_i f_{i}^4( \ \langle \! \langle  \eta_i^2 \rangle \! \rangle -1) 
\nonumber \\
&+& 2E^2
\sum_{i,j} f_{i}^2 f_{j}^2 (h_{ij})^2 
( \ \langle \! \langle  \eta_i^2 \rangle \! \rangle -1)
( \ \langle \! \langle  \eta_j^2 \rangle \! \rangle -1) 
\nonumber \\
&+&\sum_{i,k} f_{i}^2 f_{k}^2 (h_{ik}^2)^2 
( \ \langle \! \langle  \eta_i^2 \rangle \! \rangle -1)
( \ \langle \! \langle  \eta_k^2 \rangle \! \rangle -1)
\nonumber \\&+&\sum_{i,j,k} f_{i}^2 f_{k}^2 (h_{ij})^2 (h_{jk})^2
( \ \langle \! \langle  \eta_i^2 \rangle \! \rangle -1)
( \ \langle \! \langle  \eta_j^2 \rangle \! \rangle -1)
 ( \ \langle \! \langle  \eta_k^2 \rangle \! \rangle -1) \ ,
\label{sig2}
\end{eqnarray} 
where we added the suffix $E$ to clearly show that the quantities
are for an exact eigenstate. Note that we can analytically evaluate
them using $ \ \langle \! \langle  \eta_i^2 \rangle \! \rangle = a_i =
\epsilon / \mid f_i \mid$.   

In numerical study we first consider the case 
$\mid \psi \rangle = \mid \psi _{\rm E} \rangle$ on a small lattice, and then 
proceed to the case that the state is an approximate one denoted by 
$\mid \psi _{\rm A} \rangle$. In both cases we measure 
$ \ \langle \! \langle E_{{\{\eta \}}}(L) \rangle \! \rangle _{\rm smpl}$
and $ \rho_{{\{\eta \}}}^2(L)$
from $n_{\rm smpl}$ samples in the same manner as (\ref{wem1}) and
(\ref{wem2}) in the previous section, namely, 
\begin{eqnarray} 
 \ \langle \! \langle E_{{\{\eta \}}}(L) \rangle \! \rangle _{\rm smpl} &\equiv&
\frac{1}{n_{\rm smpl}} \sum_{k=1}^{n_{\rm smpl}} E_{{\{\eta \}_k}}(L) ,
\label{wem4}
\\
 \rho_{{\{\eta \}}}^2(L) & \equiv & 
 \ \langle \! \langle [E_{{\{\eta \}}}(L) ]^2 \rangle \! \rangle _{\rm smpl} 
 - \ \langle \! \langle E_{{\{\eta \}}}(L) \rangle \! \rangle _{\rm smpl} ^2 \ .
\label{wem5} 
\end{eqnarray}  
The error of 
$  E_{{\{\eta \}}}(L) $ is
evaluated by 
\begin{eqnarray}
Er(L) \equiv 2\sqrt{ \frac {\ \rho_{{\{\eta \}}}^2(L)}{n_{\rm smpl}}}. 
\label{wem6}
\end{eqnarray}

\section{Numerical Results}

In this section we report our results for the two-dimensional
Heisenberg spin model. 
The state on each site is represented by a $z$ component of the spin.
Let us first concentrate on the case of $4 \times 4$ lattice.
This lattice size is suitable for making a careful examination of the
method because we can start with the exact ground eigenstate 
$\mid \psi_{\rm E} \rangle$ and its eigenvalue $E$.   
Table II (III) shows the $L=1$ ($L=2$) results on 
$ \ \langle \! \langle E_{\rm E{\{\eta \}}}(L) \rangle \! \rangle _{\rm smpl}$
obtained with $10^4$ samples for various values of $\epsilon$. Values of 
$\sigma_{\rm E{\{\eta \}}}^2(L) $ calculated from (\ref{sig1}) or 
(\ref{sig2}) are also presented in the table.  We see that  
$\sigma_{\rm E{\{\eta \}}}^2(L) $ and 
$\rho_{\rm E{\{\eta \}}}^2(L) $ are in good agreement. 
We also see that the exact value $E= -11.2285$ lies in the range 
$\langle \! \langle E_{\rm E{\{\eta \}}}(1) \rangle \! \rangle _{\rm smpl}
  \pm  Er(1)$ except for $\epsilon = 0.05$, while 
$ \mid \langle \! \langle E_{\rm E{\{\eta \}}}(2) \rangle \! \rangle _{\rm smpl} -
E^2 \mid \ < Er(2) $ for all values of $\epsilon$.
In Table IV we present 
$ \ \langle \! \langle E_{\rm E{\{\eta \}}}(L) \rangle \! \rangle _{\rm smpl}$ 
up to $L=10$ fixing $\epsilon=0.1 $ or $\epsilon =0.01$.
We observe that the error increases as $L$ does, but its dependence on
$L$ is weak. 
The relative error for $\epsilon=0.1$ ranges from $0.11\%$($L=1$) to  
$0.48\%$($L=10$) while for $\epsilon=0.01$ it is between
$0.006\%$($L=1$) and $ 0.015\%$($L=10$). Note that  
$Er(L)$ decreases for smaller values of $\epsilon$, 
which however costs more computer resources. 
Figure 1 plots distributions of $ E_{\rm E{\{\eta \}_k}}(L)$
$(L=1,2,3,5,10)$, which shows the deviation from
the exact $\delta$ function. The abscissa is 50 times of the ratio
of each $E_{\rm E\{\eta \}_k}(L)$ $(k=1,2,\cdots, n_{\rm smpl})$ to $E_{\rm E}(L)$. 
For small values of $L$ we see that the distribution is symmetric about
the exact value $E^L$ and the peak is high. For large $L$, 
on the other hand, the peak becomes lower and the asymmetry grows; 
the data localize in the region where 
$ E_{\rm E{\{\eta \}_k}}(L) < E^L$ and the distribution has a long tail 
toward large value of $ E_{\rm E{\{\eta \}_k}}(L)$.
These features should be taken into account in the process of the error 
estimation.

How about the number of states that appear in the calculation? 
We measure $ \ \langle \! \langle  N \rangle \! \rangle _{\rm smpl}$ $before$
and $after$ we operate $M_{\{ \eta ^{(L)}\}}$ to the state 
$\hat H M_{\{ \eta ^{(L-1)}\}} \cdots \hat H 
M_{\{ \eta ^{(1)}\}} | \psi_{\rm E} \rangle $, 
which we denote 
$\langle \! \langle  N^{\rm E}_{\rm b}(L) \rangle \! \rangle _{\rm smpl}$ and 
$\langle \! \langle  N^{\rm E}_{\rm a}(L) \rangle \! \rangle _{\rm smpl}$, 
respectively.
Table V shows the results up to $L=10$.
For $\epsilon=0.1$, we observe that 
$ \ \langle \! \langle  N^{\rm E}_{\rm a}(L) \rangle \! \rangle _{\rm smpl} \ll  
\ \langle \! \langle N^{\rm E}_{\rm b}(L) \rangle \! \rangle _{\rm smpl}$ for each $L$
and $ \langle \! \langle N^{\rm E}_{\rm b}(L) \rangle \! \rangle _{\rm smpl}$ shrinks 
to almost a half of the whole vector space. 
This observation proves the merit of using the random choice matrices. 
When we set $\epsilon=0.01$, however, the results are not so
appealing because we have to deal with almost all states of the vector 
space every time we operate $\hat H$. 
We also see in the table that, as $L$ grows, 
$\langle \! \langle  N^{\rm E}_{\rm a}(L) \rangle \! \rangle _{\rm smpl}$ 
converges to a constant which depends on the value of $\epsilon$.

Let us show the results on the approximate state then. 
Here we employ $| \psi _{\rm A} \rangle = C_N \sum_{|f_i| \geq c} | i \rangle
f_{i}$ for a small constant $c$, where $C_N$ is the normalization factor. 
In other words we abandon the coefficient
$f_i$ of the exact eigenstate $| \psi _{\rm E} \rangle $ if $|f_i| < c$.
For $c = 5 \times 10^{-3}$ the number of non-zero component of
$| \psi _{\rm A} \rangle$ is 5382 and we obtain  
$\langle \psi_{\rm A} | \hat H | \psi_{\rm A} \rangle = -10.7445$, which should be compared
to $\langle \psi_{\rm E} | \hat H | \psi_{\rm E} \rangle = -11.2285$. 
Using this  $| \psi _{\rm A} \rangle$ we measure 
$ \ \langle \! \langle E_{\rm A{\{\eta \}}}(L) \rangle \! \rangle _{\rm smpl}$, 
$ \rho_{\rm A{\{\eta \}}}^2(L) $,  
$ \ \langle \! \langle  N^{\rm A}_{\rm b}(L) \rangle \! \rangle _{\rm smpl}$ and 
$ \ \langle \! \langle  N^{\rm A}_{\rm a}(L) \rangle \! \rangle _{\rm smpl}$  
up to $L=10$ with $n_{\rm smpl} = 10^4$ and compare them to exactly calculated 
$\langle \psi_{\rm A} | \hat H ^L | \psi_{\rm A} \rangle$. Errors of 
$ E_{\rm A{\{\eta \}}}(L) $'s are evaluated by 
$2\sqrt{ \ \rho_{\rm A{\{\eta \}}}^2(L)/n_{\rm smpl} }$. 
Table VI and Fig.~2 show the results with $\delta = 0.01$. 
In Table VI we see a good agreement between 
$ \ \langle \! \langle E_{\rm A{\{\eta \}}}(L) \rangle \! \rangle _{\rm smpl}$ and 
$\langle \psi_{\rm A} | \hat H ^L | \psi_{\rm A} \rangle$ for all $L$.  
We also see in Fig.~2 that for each $L$
the shape of the distribution of $E_{\rm A{\{\eta \}_k}}(L)$ is similar to
that of $E_{\rm E{\{\eta \}_k}}(L)$ in Fig.~1 although the peak is
slightly lower and the asymmetry is more outstanding.   
It should be noted that, as is expected, for larger $L$ we obtain better
evaluation for the true value $E_0$. In Table VI for example, we see   
$E_{\rm A}(1) = -10.7445 \ > \ 
- \sqrt[10]{E_{\rm A}(10)} = - \sqrt[10]{0.297136 \times 10^{11
}} = -11.1505 \ > \ E_0 = -11.2285$.

Finally we report our results on lattices larger than $4\times 4$. 
Here we use 
 \begin{eqnarray}
  E_{\{\eta\}}(L) = \langle \psi \mid \hat H 
M_{\{ \eta ^{(L)}\}} \hat H M_{\{ \eta ^{(L-1)}\}} \cdots \hat H 
M_{\{ \eta ^{(1)}\}} \mid \psi \rangle .
\end{eqnarray}
instead of (\ref{eldef}) in order to decrease the statistical errors.

In order to apply our method to large systems we have to obtain 
$| \psi _{\rm A} \rangle$ without knowing what the exact eigenstate 
$| \psi _{\rm E} \rangle$ is.
Among several attempts we made, we took the following procedure which is
based on the Suzuki-Trotter formula.
\begin{enumerate}
\item Choose one trial state $| \psi _{\rm trl}\rangle$.
\item Operate $exp(- \Delta \hat H _1) \cdot exp(- \Delta \hat H _2)$ to
$| \psi _{\rm trl}\rangle$,
 where $\hat H_1 + \hat H_2 = \hat H$ and $\Delta = 0.4$.
\item Abandon the small coefficients in 
$exp(- \Delta \hat H _1) \cdot exp(- \Delta \hat H _2)| \psi _{\rm trl} \rangle$
if necessary so that the number of non-zero components of the state is 
within the limit of our computers.  
\end{enumerate}
For the $6\times 6$ square lattice
the number of non-zero components of the $| \psi _{\rm A} \rangle$ thus obtained is 
$O(10^6)$ with the minimum absolute 
value of the non-zero coefficients $|c|_{min}= 1.14 \times 10^{-4}$.

Then we apply the on-off probability method 
setting $\epsilon = 0.0057$ and $\delta = |c|_{min}$. Table VII shows
values of 
$ \ \langle \! \langle E_{\rm A{\{\eta \}}}(L) \rangle \! \rangle _{\rm smpl}$, 
$ \ \langle \! \langle  N^{\rm A}_{\rm b}(L) \rangle \! \rangle _{\rm smpl}$ and 
$ \ \langle \! \langle  N^{\rm A}_{\rm a}(L) \rangle \! \rangle _{\rm smpl}$  
up to $L=5$ with $5.5 \times 10^3$ samples. 
The memory we need here is 356 MBytes including the hash table 
and the CPU time consumed to one sample is 1100 seconds by a  
Pentium III machine. 

Now we have obtained reliable estimations of 
$ E_{\rm A}(L)  = \langle \psi_{\rm A} \mid \hat H ^L \mid \psi_{\rm A} \rangle$
$(L = 1, 2, \cdots, 5)$. How can we estimate physical quantities, $E_0$
for instance, using them? 
We see that $\sqrt[5]{\langle \! \langle E_{\rm A\{\eta\}}(5) \rangle \!
\rangle_{\rm smpl}}= -23.68683 \pm 0.00083$ is lower than    
$ E_{\rm A}(1) = -23.56093$. But, because $L$ is not large enough, even 
$\sqrt[5]{\langle \! \langle E_{\rm A\{\eta\}}(5) \rangle \! \rangle_{\rm smpl}}$
provides only a poor upper bound for the
true value $E_0 = -24.4394$ which is obtained in ref.(5) 
by the exact diagonalization. 
One better way for the estimation would be to calculate the lowest eigenvalue 
$E$ starting from the basis $\{| \psi_{\rm A} \rangle,  \hat H | \psi_{\rm A} \rangle, 
\hat H ^2| \psi_{\rm A} \rangle \}$. It is easy to obtain the 
orthonormalized basis 
$\{ | \psi_0 \rangle, | \psi_1 \rangle, | \psi_2 \rangle \}$ 
from  $\{| \psi_{\rm A} \rangle,  \hat H | \psi_{\rm A} \rangle, 
\hat H ^2| \psi_{\rm A} \rangle \}$ and to
calculate the Hamiltonian matrix elements which are the functions of  
$E_{\rm A}(L)$ $(L \le 5)$. It, however, turned out to be ineffective too; 
the obtained value is $E=-23.88$, which is still too far from $E_0$. 
So we need additional assumptions to make a better estimate of the eigenvalue. 
Here, based on the low energy property described by the linear
spin wave theory or more sophisticated models\cite{lsw}, 
we empirically assume 
\begin{eqnarray} 
\langle \psi_{\rm A} \mid \hat H ^L \mid \psi_{\rm A} \rangle &=& 
q_0 E^L + q_1 \int_0^1 (xE)^L \cdot  x^\alpha dx \nonumber \\
&=& E^L (q_0 + \frac{q_1}{L +\alpha +1}) \equiv F(L,E,q_0,q_1,\alpha) \ ,
\label{empias}
\end{eqnarray}  
where $q_0$, $q_1$ and $\alpha$ are free parameters which should be 
determined together with $E$ by the fit. Note that $q_0 + q_1/(\alpha+1) = 1$ 
should hold due to the normalization of $| \psi_{\rm A} \rangle$.
We look for the minimum of the difference $D$ with  $L_{\rm max}=5$,
\begin{eqnarray*}
D \equiv \sum_{L=1}^{L_{\rm max}}
\left[  1- \frac{ \ \langle \! \langle E_{\rm A{\{\eta \}}}(L) 
\rangle \! \rangle _{\rm smpl}}{F(L,E,q_0,q_1,\alpha)} \right]^2,
\end{eqnarray*} 
changing these four parameters and accept them if 
$\mid q_0 + q_1/(\alpha+1) -1 \mid \le 0.003$ is fulfilled and 
$D$ is less than the sum of the relative errors, which amounts 
\begin{eqnarray*}
 \sum_{L=1}^{L_{\rm max}} \left[
\frac{Er(L)}{ \ \langle \! \langle E_{\rm A{\{\eta \}}}(L) \rangle \!
\rangle _{\rm smpl}}\right] ^2 \ = 8.5 \times 10^{-9} \ . 
\end{eqnarray*} 
By this fit we obtain $-24.32 \le E \le -24.13$, which indicates that the
evaluation is useful. 

It is possible to advance to $8 \times 8 $ lattice in the same manner,
where the number of non-zero components of $| \psi _{\rm A} \rangle$ is
$O(10^7)$ and $|c|_{\rm min}= 3.28 \times 10^{-5}$. The results with
$n_{\rm smpl} = 1.1 \times 10^3$, $\epsilon = 0.01$ and $\delta =
|c|_{\rm min}$ 
are shown in Table VIII. Note that 
$\sqrt[5]{\langle \! \langle E_{\rm A\{\eta\}}(5) \rangle \! \rangle
_{\rm smpl}} = -41.03348 \pm 0.03817$ is lower than $ E_{\rm A}(1)=-40.74998 $.
By the fitting using (\ref{empias}) we obtain an estimation 
$-43.5 \le E \le -41.1$, which includes $-43.103$, the value also reported in 
ref.(5). 
It is large sample errors, which would be improved by increasing
$n_{\rm smpl}$, that cause the wide range of the evaluated $E$ here. 
  
\section{Summary and Comments}

In this paper we have suggested an alternative Monte Carlo approach
which, using the stochastic variables following the on-off probability 
function, does not rely on random walks nor importance samplings.
Our purpose is to show the way to study large size systems without 
keeping the full vector space. 
We apply this method to the two-dimensional Heisenberg spin one-half 
model, which was chosen as a familiar example, and 
carry out numerical evaluations for lattices up to $8\times 8$. 
The results show that we can obtain reliable numerical data on 
expectation values of the power of its Hamiltonian with reasonable 
computer resources.

A few remarks are in order.
  
One merit of this method is that it enables us to numerically evaluate 
expectation values in various quantum systems using relatively small 
portion of the whole vector space, which therefore enlarge the size of
the systems we study.
Another merit is that the method can be applied to systems which 
have negative sign problem. Actually a work on the 
Shastry-Sutherland model\cite{ss,ov} near the critical point as well as
on the quantum spin model on a triangular lattice is in
progress\cite{newmunes}.  

More studies should be necessary on two issues of the method.
One of them is how one can effectively generate the approximate state 
$\mid \psi _{\rm A} \rangle$. Although we can improve our estimation by
increasing the number of samples $n_{\rm smpl}$, it is desirable, in order
to perform efficient estimations, to employ a good 
$\mid \psi _{\rm A} \rangle$ for the following reasons. 
One reason is that with the better 
$\mid \psi _{\rm A} \rangle$ the sample error will be smaller.
Another reason is that with the better $\mid \psi _{\rm A} \rangle$ 
the smaller $L$ will be enough to obtain reliable estimates.
In addition to the very empirical way we adopted to obtain  
$\mid \psi _{\rm A} \rangle$ in this paper, a systematic way which 
makes use of the Lanczos method 
in small vector spaces are under investigation. There we start with a
trial state $ \mid \psi_{\rm trl} \rangle$, calculate the matrix elements of the 
Hamiltonian $\hat H$, diagonalize the matrix and define 
$\mid \psi _{\rm A} \rangle$ as
its lowest energy eigenstate.  The basis for the matrix elements is 
$\{ \mid \psi_{\rm trl} \rangle, \hat H \mid \psi_{\rm trl} \rangle, 
\hat H^2 \mid \psi_{\rm trl} \rangle, \cdots, \hat H^p \mid \psi_{\rm trl} \rangle \}$
with a small value of $p$, which we numerically calculate without 
any approximation.
Another important problem is to find a sophisticated way to extract physical 
quantities from the expectation values. The results of several fits we
tried seem to suggest there are some shortages in our simple
assumptions.
Both of the issues, which might be largely model-dependent, should be 
investigated in the future work.  

The on-off probability method could be applied to any problem  
which is mathematically related with the Markov process. In the Markov
process, which  can be described by repeated operation of a probability
matrix $M_{\rm p}$ to an initial state $\mbox{\boldmath $v$ }$, rapid increase of
states would occur. If we can replace $(M_{\rm p})^L \ \mbox{\boldmath $v$ }$ with 
$M_{\{ \eta ^{(L)}\}}  M_{\rm p}  \cdots M_{\{ \eta ^{(1)}\}}  M_{\rm p} \ \mbox{\boldmath $v$ }$
the number of states we should consider becomes much smaller. 
We hope our method is helpful in many fields of science. 



\newpage
\vskip 1cm
\begin{table}[h]
\centering
\begin{tabular}{cccccc} \hline
$\epsilon$  &
$\langle \! \langle  S_{\{ \eta \}} \rangle \! \rangle _{\rm smpl}  $
   & $\sigma_{\rm S}^2  $ &  $\rho_{\rm S}^2  $ 
& $\langle \! \langle  N \rangle \! \rangle  $ &
 $\langle \! \langle  N \rangle \! \rangle _{\rm smpl} $ 
  \\ \hline
 0.3 & 73.572$\pm $0.091 &  21.07 &  20.81     &   245.24 & 245.3  \\ \hline
 0.2 & 73.582$\pm $0.073 &  13.76 & 13.62     &  366.99 & 367.0  \\ \hline
 0.1 & 73.558$\pm $0.050&  6.46 &  6.34     & 731.97 &  731.8  \\ \hline
 0.05 & 73.591$\pm $0.033 & 2.84&  2.77     & 1451.72 &  1452.1  \\ \hline
\end{tabular}
\caption{
Results on $S=\sum |c_i|$ calculated by the on-off probability
method, where $c_i $ denotes the coefficient of state $ |i \rangle$ 
for the exact ground state of the $4\times 4$ Heisenberg quantum spin system. 
The number of samples, $n_{\rm smpl}$, is $10^4$.
The exact value of $S$ is 73.57224.
The values of $\sigma_{\rm S}^2 $ and
$\langle \! \langle  N \rangle \! \rangle  $ calculated by
 (\ref{sigdef}) and (\ref{ndef}) are also shown in the table. }
\end{table}

\begin{table}[h]
\centering
\begin{tabular}{cccc} \hline
$\epsilon$  &
$\langle \! \langle E_{\rm E{\{\eta \}}}(1) \rangle \! \rangle _{\rm smpl} $
 & $\sigma_{\rm E{\{\eta \}}}^2(1) $ & $\rho_{\rm E{\{\eta \}}}^2(1)$  \\ \hline 
 0.20 & $-$11.2128$\pm $0.0215&1.1703 &1.168 \\ \hline
 0.15 & $-$11.2198$\pm $0.0170&0.7345 &0.738 \\ \hline
 0.10 & $-$11.2285$\pm $0.0122&0.3606 &0.374 \\ \hline
 0.08 & $-$11.2212$\pm $0.0095&0.2283 &0.228 \\ \hline
 0.05 & $-$11.2230$\pm $0.0052&0.07338 &0.0719 \\ \hline
 0.03 & $-$11.2288$\pm $0.0033&0.02731 &0.0274 \\ \hline
 0.02 & $-$11.2266$\pm $0.0021&0.01037 &0.0106 \\ \hline
 0.01 & $-$11.2283$\pm $0.0007&0.00120 &0.0012 \\ \hline
\end{tabular}
\caption{
Results on $E_{\rm E}(1) = \langle \psi_{\rm E} | \hat H | \psi_{\rm E} \rangle$
obtained for the $4\times 4$
Heisenberg quantum spin system by the on-off probability method
with $n_{\rm smpl} = 10^4$,
where $| \psi _{\rm E}  \rangle$ is the exact eigenstate of the system.
The exact value is $E= -11.2285$.
    }
\end{table}
\begin{table}[h]
\centering
\begin{tabular}{cccc} \hline
$\epsilon$  
&$\langle \! \langle E_{\rm E{\{\eta \}}}(2) \rangle \! \rangle _{\rm smpl}$
& $\sigma_{\rm E{\{\eta \}}}^2(2)$ &  $\rho_{\rm E{\{\eta \}}}^2(2) $  \\ \hline
 0.20 &  126.200$\pm $0.344&297.23 &296.2 \\ \hline
 0.15 &  126.006$\pm $0.262&170.68 &171.8 \\ \hline
 0.10 &  125.988$\pm $0.175& 77.04 & 76.7 \\ \hline
 0.08 &  126.006$\pm $0.138& 47.40 & 48.2 \\ \hline
 0.05 &  126.062$\pm $0.013& 14.78 & 14.9 \\ \hline
 0.03 &  126.088$\pm $0.046&  5.332 &  5.30 \\ \hline
 0.02 &  126.079$\pm $0.028&  1.997 &  2.01 \\ \hline
 0.01 &  126.071$\pm $0.009&  0.228 &  0.23 \\ \hline
\end{tabular}
\caption{
Results on $E_{\rm E}(2) = \langle \psi_{\rm E} | \hat H^2 | \psi_{\rm E} \rangle$ 
obtained for the $4\times 4$ Heisenberg quantum spin system 
by the on-off probability method with $n_{\rm smpl}=10^4$.
The exact value is $E^2=  126.0788$.}
\end{table}
\begin{table}[h]
\centering
\begin{tabular}{rrlrlrl} \hline 
\multicolumn{1}{c}{ } &  \multicolumn{2}{c}{ }
&\multicolumn{2}{c}{$ \epsilon =0.1$}
  &\multicolumn{2}{c}{$ \epsilon =0.01$} 
\\ \cline{4-7}
\multicolumn{1}{c}{$L$}
& \multicolumn{2}{c}{$E_{\rm E}(L)$} & \multicolumn{4}{c}
{$\langle \! \langle E_{\rm E{\{\eta \}}}(L) \rangle \! \rangle _{\rm smpl}$} 
 \\ \hline
  1  & $-$0.112285 &$ \times  10^{ 2}$ & $-$(0.11228 $\pm $ 0.00012) & 
$ \times  10^{ 2}$ & $-$(0.112284 $\pm $ 0.000007) &
$ \times  10^{ 2}$ \\ \hline
  2  &  0.126079 & $ \times  10^{ 3}$ & (0.12599 $\pm $ 0.00017) &
 $ \times  10^{ 3}$ & (0.126078 $\pm $ 0.000009)&
 $ \times  10^{ 3}$ \\ \hline
  3  & $-$0.141567 & $ \times  10^{ 4}$ & $-$(0.14163 $\pm $ 0.00023) & 
$ \times  10^{ 4}$ & $-$(0.141567 $\pm $ 0.000012)&
 $ \times  10^{ 4}$ \\ \hline
  4  &  0.158959 & $ \times  10^{ 5}$ & (0.15900 $\pm $ 0.00031)&
 $ \times  10^{ 5}$ & (0.158952 $\pm $ 0.000016)&
 $ \times  10^{ 5}$ \\ \hline
  5  & $-$0.178486 & $ \times  10^{ 6}$ & $-$(0.17846 $\pm $ 0.00040)&
 $ \times  10^{ 6}$ & $-$(0.178481 $\pm $ 0.000020)&
 $ \times  10^{ 6}$ \\ \hline
  6  &  0.200413 & $ \times  10^{ 7}$ & (0.20023 $\pm $ 0.00051) &
 $ \times  10^{ 7}$ & (0.200408 $\pm $ 0.000023) &
 $ \times  10^{ 7}$ \\ \hline
  7  & $-$0.225034 & $ \times  10^{ 8}$ & $-$(0.22498 $\pm $ 0.00067)& 
 $ \times  10^{ 8}$ & $-$(0.225029 $\pm $ 0.000028) & 
$ \times  10^{ 8}$ \\ \hline
  8  &  0.252679 & $ \times  10^{ 9}$ & (0.25218 $\pm $ 0.00081)&
 $ \times  10^{ 9}$ & (0.252684 $\pm $ 0.000034) & 
$ \times  10^{ 9}$ \\ \hline
  9  & $-$0.283720 & $ \times  10^{10}$ & $-$(0.2835 $\pm $ 0.0010) & 
$ \times  10^{10}$ & $-$(0.283715 $\pm $ 0.000040)&
 $ \times  10^{10}$ \\ \hline
 10  &  0.318574 & $ \times  10^{11}$ & (0.3182 $\pm $ 0.0015) & 
$ \times  10^{11}$ & (0.318591 $\pm $ 0.000047) & $ \times  10^{11}$ \\ \hline
\end{tabular}
\caption{
Results on $E_{\rm E}(L) = \langle \psi_{\rm E} | \hat H ^L| \psi_{\rm E} \rangle $ 
$(L=1,2, \cdots, 10)$ obtained for the $4\times 4$ Heisenberg quantum 
spin system by the on-off probability method with $10^4$ samples. We
present the data for $\epsilon = 0.1$ and $\epsilon = 0.01$ together with the
exact values of $E_{\rm E}(L)$.}
\end{table}
\begin{table}[h]
\centering
\begin{tabular}{rcccc} \hline
 { }&\multicolumn{2}{c}{$ \epsilon =0.1$}
  &\multicolumn{2}{c}{$ \epsilon =0.01$}   \\ \cline{2-5}
\multicolumn{1}{c}{$L$}
&$ \ \langle \! \langle N^{\rm E}_{\rm b}(L) \rangle \! \rangle _{\rm smpl}$   & $ \ \langle \! \langle  N^{\rm E}_{\rm a}(L) \rangle \! \rangle _{\rm smpl} $  & 
$ \ \langle \! \langle N^{\rm E}_{\rm b}(L) \rangle \! \rangle _{\rm smpl}$ & $ \ \langle \! \langle  N^{\rm E}_{\rm a}(L) \rangle \! \rangle _{\rm smpl} $  \\ \hline
 1 & 12870  & 731.7 & 12870 &    6249.41 \\ \hline
 2&7959.5 &   564.7 & 12797.06 &   6242.70 \\ \hline
 3 &    6664.4&   515.6 &  12796.22&   6242.22 \\ \hline
4 &   6238.6 &   498.5&  12796.22 &   6242.85 \\ \hline
5&   6083.2 &   491.9& 12796.34 &    6242.79 \\ \hline
6 &   6022.0&   488.9&  12796.25 &    6243.20 \\ \hline
7&   5995.0 &   487.9& 12796.27 &   6242.38 \\ \hline
8 &  5985.5 &   487.3 &  12795.97 &  6243.45 \\ \hline
9 &   5981.4 & 487.2& 12796.42 &   6242.71\\ \hline
10 &  5980.1 &  487.5& 12796.21 &   6243.33 \\ \hline
\end{tabular}
\caption{
Numbers of non-zero coefficients $before$ and $after$ operating the 
random choice matrix $M_{\{ \eta ^{(L)}\}}$ to the state 
$\hat H M_{\{ \eta ^{(L-1)}\}} \cdots \hat H 
M_{\{ \eta ^{(1)}\}} | \psi_{\rm E} \rangle $, where $\hat H$ and 
$| \psi_{\rm E} \rangle$ 
denote the Hamiltonian and the exact ground state of the 
$4\times 4$ Heisenberg quantum spin system, respectively.
The number of samples is $10^4$. }
\end{table}
\begin{table}[h]
\centering
\begin{tabular}{rrlrlcc} \hline
\multicolumn{1}{c}{$L$}
& \multicolumn{2}{c}{$E_{\rm A}(L)$} & \multicolumn{2}{c}
{$ \ \langle \! \langle  E_{\rm A{\{\eta\}}}(L)\rangle \! \rangle _{\rm smpl}$}
 & $\langle \! \langle  N^{\rm A}_{\rm b}(L) \rangle \! \rangle _{\rm smpl}$
 &  $\langle \! \langle  N^{\rm A}_{\rm a}(L) \rangle \! \rangle _{\rm smpl}$  \\ \hline
  1  & $-$0.107445& $ \times  10^{ 2}$ & $-$(0.10750 $\pm $ 0.00012) & $ \times  10^{ 2}$ & 5382.0 &540.97  \\ \hline
  2  &  0.119302& $ \times  10^{ 3}$ & (0.11937 $\pm $ 0.00017) & $ \times  10^{ 3}$ & 6342.2 &466.83  \\ \hline
  3  & $-$0.133162& $ \times  10^{ 4}$ & $-$(0.13327 $\pm $ 0.00025) & $ \times  10^{ 4}$ & 5739.5 &445.39  \\ \hline
  4  &  0.149066& $ \times  10^{ 5}$ & (0.14936 $\pm $ 0.00038) & $ \times  10^{ 5}$ & 5550.9 &438.58  \\ \hline
  5  & $-$0.167056& $ \times  10^{ 6}$ & $-$(0.16720 $\pm $ 0.00062) & $ \times  10^{ 6}$ & 5488.3 &436.49  \\ \hline
  6  &  0.187354& $ \times  10^{ 7}$ & (0.1876 $\pm $ 0.0012) & $ \times  10^{ 7}$ & 5468.6 &435.69  \\ \hline
  7  & $-$0.210193& $ \times  10^{ 8}$ & $-$(0.2121 $\pm $ 0.0028) & $ \times  10^{ 8}$ & 5462.0 &435.78  \\ \hline
  8  &  0.235870& $ \times  10^{ 9}$ & (0.2368 $\pm $ 0.0034) & $ \times  10^{ 9}$ & 5461.7 &435.47  \\ \hline
  9  & $-$0.264725& $ \times  10^{10}$ & $-$(0.2647 $\pm $ 0.0033) & $ \times  10^{10}$ & 5459.3 &435.24  \\ \hline
 10  &  0.297136& $ \times  10^{11}$ & (0.2958 $\pm $ 0.0049) & $ \times  10^{11}$ & 5456.6 &435.58  \\ \hline
\end{tabular}
\caption{
Results on $E_{\rm A}(L) \equiv \langle \psi_{\rm A} | H^L | \psi_{\rm A} \rangle$ for 
the $4\times 4$ system obtained from $10^4$ samples with $\epsilon =0.1$
and $\delta = 0.01$.  
We also present the exact values of $E_{\rm A}(L)$ in the second column 
for comparison.
    }
\end{table}
\begin{table}[h]
\centering
\begin{tabular}{rrlcc} \hline
\multicolumn{1}{c}{$L$}
& \multicolumn{2}{c}
{$ \ \langle \! \langle  E_{\rm A{\{\eta\}}}(L)\rangle \! \rangle _{\rm smpl}$}
   & $\langle \! \langle  N^{\rm A}_{\rm b}(L) \rangle \! \rangle _{\rm smpl}$ 
&  $\langle \! \langle  N^{\rm A}_{\rm a}(L) \rangle \! \rangle _{\rm smpl}$   \\ \hline 
  1  & $-$(0.235605 $\pm $ 0.000005) & $ \times  10^{ 2}$ & 1173554 & 55121.7  \\ \hline
  2  &(0.557681 $\pm $ 0.000023) & $ \times  10^{ 3}$ & 2159459.8 & 72578.1  \\ \hline
  3  & $-$(0.132179 $\pm $ 0.000008) & $ \times  10^{ 5}$ & 2952153.5 & 88743.8  \\ \hline
  4  &(0.313746 $\pm $ 0.000029)  &$ \times  10^{ 6}$ & 3677982.2 &103182.8  \\ \hline
  5  & $-$(0.74565 $\pm $ 0.00013)& $ \times  10^{ 7}$ & 4301258.0 &115698.3  \\ \hline
\end{tabular}
\caption{
Results on $E_{\rm A}(L)$ for the $6\times 6$ system obtained 
from $5.5 \times 10^3$ samples with $\epsilon =0.0057$. 
The exact value of 
$E_{\rm A}(1) = \langle \psi _{\rm A} \mid \hat H \mid \psi _{\rm A} \rangle $ is 
calculated to be $-23.56093 $.
    }
\end{table}
\begin{table}[h]
\centering
\begin{tabular}{rrlcc} \hline
\multicolumn{1}{c}{$L$}
& \multicolumn{2}{c}{
$ \ \langle \! \langle  E_{\rm A{\{\eta\}}}(L)\rangle \! \rangle _{\rm smpl} $}  
 & $\langle \! \langle  N^{\rm A}_{\rm b}(L) \rangle \! \rangle _{\rm smpl}$
&  $\langle \! \langle  N^{\rm A}_{\rm a}(L) \rangle \! \rangle _{\rm smpl}$  \\ \hline 
  1  &$-$(0.407475 $\pm $ 0.000075) & $ \times  10^{ 2}$ & 9549922 & 96800.9  \\ \hline
  2  &(0.167062 $\pm $ 0.000054) & $ \times  10^{ 4}$ & 9572198.2 & 73091.7  \\ \hline
  3  & $-$(0.68731 $\pm $ 0.00047) & $ \times  10^{ 5}$ & 7155664.2 & 63898.6  \\ \hline
  4  &(0.28338 $\pm $ 0.00090) & $ \times  10^{ 7}$ & 6203627.8 & 60498.0  \\ \hline
  5  & $-$(0.11633 $\pm $ 0.00054) & $ \times  10^{ 9}$ & 5857166.8 & 59281.9  \\ \hline
\end{tabular}
\caption{
Results on $E_{\rm A}(L)$ for the $8\times 8$ system obtained 
from $1.1 \times 10^3$ samples with $\epsilon =0.01$. 
The exact value of $E_{\rm A}(1)$ is $-40.74998 $. 
    }
\end{table}
\begin{figure}[ht]
\begin{center}
\scalebox{0.5}{\includegraphics{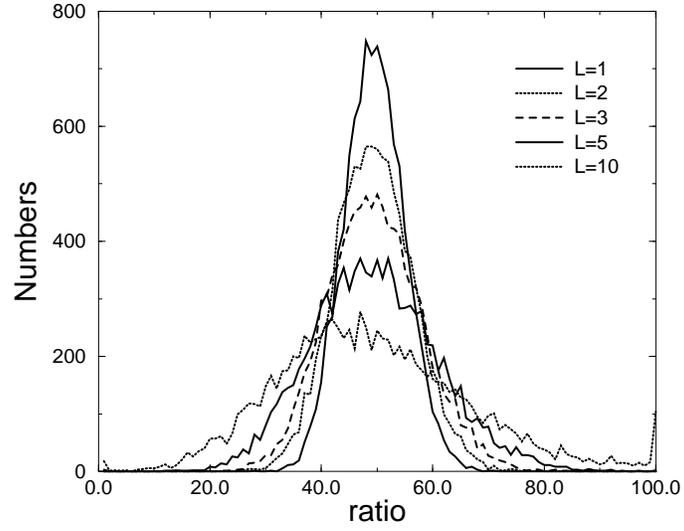}}
\caption{
Distributions of $ E_{\rm E \{ \eta \}_k}(L)$ $(k=1,2, \cdots, 10^4)$
on the $4\times 4$ lattice for several values of $L$. Here the $ratio$
in the horizontal axis is defined as 50 times of $E_{\rm E\{\eta \}_k}(L)/E_{\rm E}(L)$.
}
\end{center}
\end{figure}
\begin{figure}[htp]
\begin{center}
\scalebox{0.5}{\includegraphics{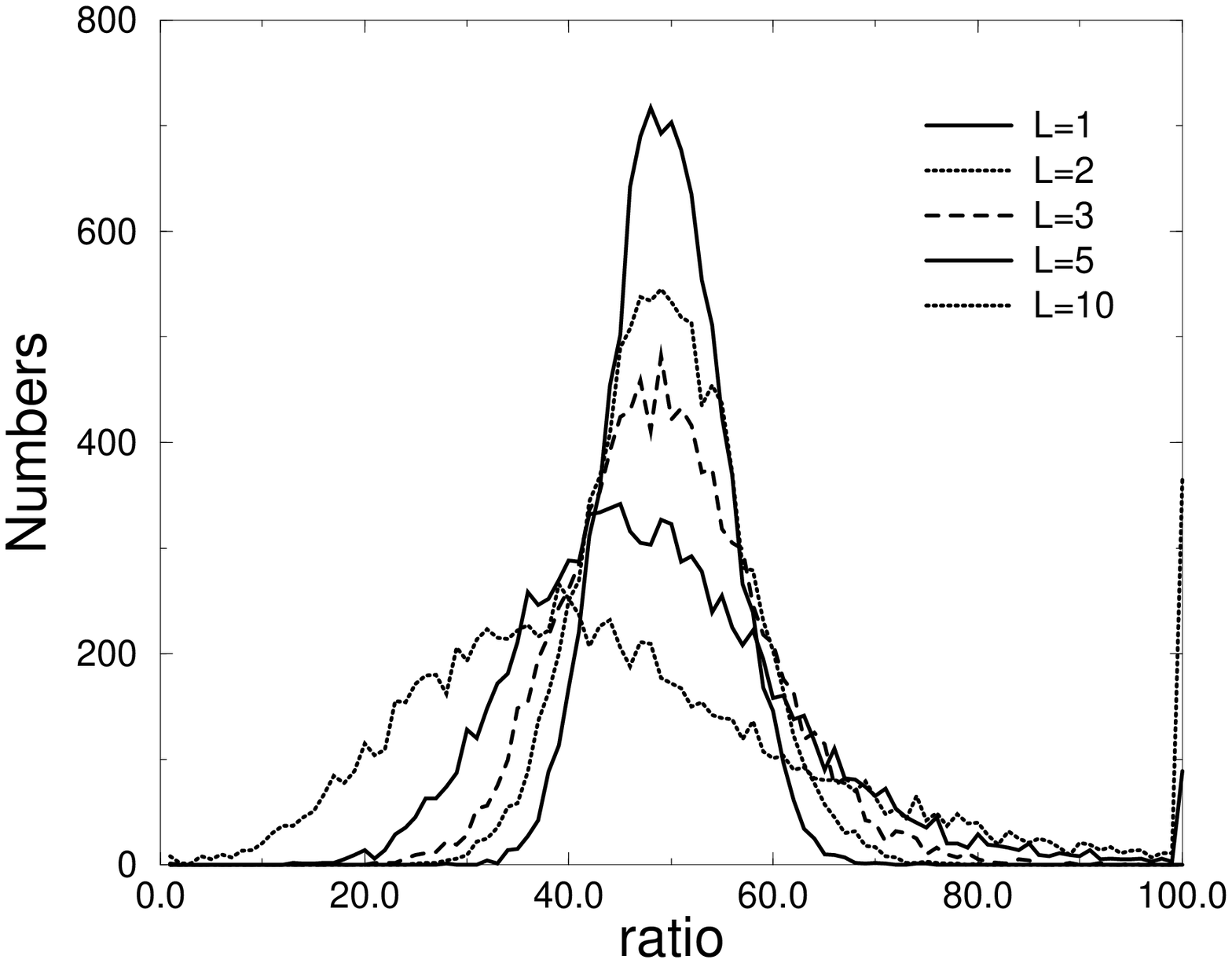}}
\caption{Distributions of $E_{\rm A \{ \eta \}_k}(L)$ $(k=1,2, \cdots, 10^4)$
on the $4\times 4$ lattice for several values of $L$. The $ratio$
is 50 times of $E_{\rm A\{\eta \}_k}(L)/E_{\rm A}(L)$.}
\end{center}
\end{figure}


\begin{thebibliography}{99}
\bibitem{foot1}
It holds for $any$ probability function.
If the probability function is Gaussian the confidence level rises to
95.4 \%.
\end{thebibliography}
\end{document}